# APPLICABILITY OF HULTHÉN-HELLMANN POTENTIAL TO PREDICT THE MASS-SPECTRA OF HEAVY MESONS VIA SERIES EXPANSION METHOD

[1,2]Inyang, E. P., [2]Ntibi, J. E., [1]Ibanga, E. A. and [2]William, E. S.

[1]Department of Physics, National Open University of Nigeria, Jabi, Abuja, Nigeria

[2]Theoretical Physics Group, Department of Physics, University of Calabar, P.M.B 1115 Calabar Nigeria

Corresponding author email: etidophysics@gmail.com OR einyang@noun.edu.ng

**ABSTRACT**

We adopt Hulthén plus Hellmann potential as the quark-antiquark interaction potential for predicting the mass spectra of heavy mesons. The adopted potential was made to be temperature-dependent by replacing the screening parameter with Debye mass $(m_D(T))$. The radial Schrödinger equation was analytically solved using the series expansion method and energy eigenvalues were obtained. The energy eigenvalues is used to predict the mass spectra of heavy mesons such as charmonium $(c\bar{c})$ and bottomonium $(b\bar{b})$. Four special cases were considered when some of the potential parameters were set to zero, resulting in Hellmann potential, Yukawa potential, Coulomb potential, and Hulthén potential, respectively. The present potential provides satisfying results in comparison with experimental data and the work of other researchers with a maximum error of 0.034 GeV.

## INTRODUCTION

The solution of the Schrödinger equation (SE) for a physical system in quantum mechanics is of great importance, because the knowledge of Eigen energy and wave function contains all possible information about the physical properties of a system under study (Inyang et al., 2021; Allosh et al., 2021). The study of behavior of several physical problems in physics requires us to solve the non-relativistic or relativistic equation. A good description of many features of these problems can be obtained using non-relativistic models that is the quark-antiquark strong interaction is described by a phenomenological potential (Abu-shady et al., 2021). Heavy mesons have turned out to provide extremely useful probes for the deconfined state of matter because the force between a heavy quark and anti-quark is weakened due to the presence of gluons which lead to the dissociation of it bound states (Vega and Flores, 2016). The heavy mesons and their interaction are well described by the SE (Prasanth et al., 2020; Inyang et al., 2021).

The solution of the spectral problem for the SE with spherically symmetric potentials is of major concern in describing the spectra of heavy mesons (Rani et al., 2018). Potential models offer a rather good description of the mass spectra of heavy mesons such as bottomonium, and charmonium (Mansour and Gamal, 2018). In predicting the mass spectra of heavy mesons, confining-type potentials are generally used. The holding potential is the Cornell potential with two terms one of which is responsible for the Coulomb interaction of the quarks and the other corresponds to a confining term (Al-Oun et al., 2015). In past, this type of potential has been studied by many researchers using different techniques (Al-Jamel 2019; Abu-Shady,2016; Ciftci, and Kisoglu, 2018;Bayrak et al.,2006;Hall and Saad,2015). The confining potentials may be in different forms depending upon the interaction of the particles within the system. Harmonic oscillator and hydrogen atom are the two potentials which solutions to the SE are found exactly. On the other hand, to obtain the approximate solutions, some techniques are employed. Example of such techniques include, asymptotic iteration method (AIM) (Okorie et al.,2021),Laplace transformation method (Abu-Shady and Khokha,2018), super symmetric quantum mechanics (SUSYQM) (Abu-Shady and Ikot, 2019), the Nikiforov-Uvarov (NU) method (Okoi et al.,2020; Inyang et al., 2021;Edet et al., 2020;Inyang et al.,2020;Inyang et al., 2021; Edet et al., 2020;William et al., 2020;Inyang et al.,2021; Okon et al.,2016;Abu-Shady et al.,2019; Inyang et al., Omugbe, 2020;Thompson et al.,2021;Abu-Shady,2015;Akpan et al.,2021;Inyang et al.,2021) ,the Nikiforov-Uvarov Functional Analysis (NUFA) method (Ikot et al., 2021), the series expansion method (SEM) (Inyang et al.,2021;Abu-Shady and Fath-Allah, 2019;Inyang et al.,2021; Ibekwe et al., 2021),analytical exact iterative method(AEIM)(Khokha et al., 2016),WKB approximation method (Omugbe et al., 2020;Omugbe et al.,2021;Omugbe et al.,2022;), Exact quantization rule (EQR) (Qiang et al.,2008;Inyang et al.,2020) and others (Mutuk, 2019) .The Hulthén potential, (1942) is a short-range potential that behaves like a Coulomb potential for small values of $r$ and decreases exponentially for large values of $r$ . It has been used in many branches of physics, such as nuclear and particle physics, atomic physics, solid-state physics, and chemical physics (Nwabuzor et al.,2021).

The Hellmann potential,(1935) which is a superposition of an attraction Coulomb potential and a Yukawa potential has been studied extensively by many authors in obtaining the energy of the bound state in atomic, nuclear, and particle physics (Rai and Rathaud,2015;Nasser et al., 2014).

Recently, there has been great interest in combining two potentials in both the relativistic and non-relativistic regime (William et al., 2020). The essence of combining two or more physical potential models is to have a wider range of applications. Hence, in the present work, we aim at solving the SE with the combination of Hulthén and Hellmann potential (HHP) analytically using series expansion method and apply the results to predict the mass spectra of heavy mesons such as bottomonium and charmonium, in which the quarks are considered as spinless particles for easiness. The adopted potential is of the form (William et al., 2020)

$$V(r) = -\frac{A_0 e^{-\alpha r}}{1-e^{-\alpha r}} - \frac{A_1}{r} + \frac{A_2 e^{-\alpha r}}{r} \qquad (1)$$

where $A_0, A_1$ and $A_2$ are potential strength parameters and $\alpha$ is the screening parameter. To make Eq. (1) temperature dependent, the screening parameter is replaced with Debye mass $(m_D(T))$ which vanishes at T→ 0 and we have,

$$V(r,T) = -\frac{A_0 e^{-m_D(T)r}}{1-e^{-m_D(T)r}} - \frac{A_1}{r} - \frac{A_2 e^{-m_D(T)r}}{r} \qquad (2)$$

The expansion of the exponential terms in Eq. (2) (up to order three, in order to model the potential to interact in the quark-antiquark system) yields,

$$\frac{e^{-m_D(T)r}}{r} = \frac{1}{r} - m_D(T) + \frac{m_D^2(T)r}{2} - \frac{m_D^3(T)r^2}{6} + \ldots \quad (3)$$

$$\frac{e^{-m_D(T)r}}{1-e^{-m_D(T)r}} = \frac{1}{m_D(T)r} - \frac{1}{2} + \frac{m_D(T)r}{12} + \ldots \quad (4)$$

The substitution of Eqs. (3) and (4) into Eq. (2) gave Eq. (5)

$$V(r,T) = -\frac{\beta_0}{r} + \beta_1 r - \beta_2 r^2 + \beta_3 \quad (5)$$

where

$$\left.\begin{array}{l} -\beta_0 = -A_1 + A_2 - \dfrac{A_0}{m_D(T)}, \quad \beta_1 = \dfrac{A_2 m_D^2(T)}{2} - \dfrac{A_0 m_D(T)}{12} \\ \beta_2 = \dfrac{A_2 m_D^3(T)}{6}, \quad \beta_3 = \dfrac{A_0}{2} - A_2 m_D(T) \end{array}\right\} \quad (6)$$

The first term in Eq. (5) is the Coulomb potential that describes the short distance between quarks, while the second term is a linear term for confinement feature.

## SOLUTIONS OF THE SCHRÖDINGER EQUATION WITH HULTHÉN PLUS HELLMANN POTENTIAL

We consider the radial SE of the form (Ibekwe et al., 2021)

$$\frac{d^2 R(r)}{dr^2} + \frac{2}{r}\frac{dR(r)}{dr} + \left[\frac{2\mu}{\hbar^2}(E_{nl} - V(r)) - \frac{l(l+1)}{r^2}\right]R(r) = 0 \quad (7)$$

Where $l$ is angular quantum number taking the values 0,1,2,3,4…, $\mu$ is the reduced mass for the heavy mesons, $r$ is the inter nuclear separation and $E_{nl}$ denotes the energy eigenvalues of the system.

The substitution of Eq. (5) into Eq. (7) gives

$$\frac{d^2 R(r)}{dr^2} + \frac{2}{r}\frac{dR(r)}{dr} + \left[\varepsilon + \frac{P}{r} - Qr + Sr^2 - \frac{L(L+1)}{r^2}\right]R(r) = 0 \quad (8)$$

where

$$\left.\begin{array}{l}\varepsilon = \dfrac{2\mu}{\hbar^2}(E_{nl} - \beta_3), \mathrm{P} = \dfrac{2\mu\beta_0}{\hbar^2} \\ Q = \dfrac{2\mu\beta_1}{\hbar^2}, \mathrm{S} = \dfrac{2\mu\beta_2}{\hbar^2}\end{array}\right\} \qquad (9)$$

$$L(L+1) = l(l+1) \qquad (10)$$

From Eq. (10)

$$L = -\dfrac{1}{2} + \dfrac{1}{2}\sqrt{(2l+1)^2} \qquad (11)$$

Now make an anzats wave function (Rani et al., 2018)

$$R(r) = e^{-\alpha r^2 - \beta r} F(r) \qquad (12)$$

where $\alpha$ and $\beta$ are positive constants whose values are to be determined in terms of potential parameters.

Differentiating Eq. (12) twice gives

$$R'(\mathrm{r}) = F'(\mathrm{r})e^{-\alpha r^2 - \beta r} + F(r)(-2\alpha r - \beta)e^{-\alpha r^2 - \beta r} \qquad (13)$$

$$\begin{array}{l}R''(\mathrm{r}) = F''(\mathrm{r})e^{-\alpha r^2 - \beta r} + F'(r)(-2\alpha r - \beta)e^{-\alpha r^2 - \beta r} \\ + \left[(-2\alpha) + (-2\alpha r - \beta)(-2\alpha r - \beta)\right] F(r)e^{-\alpha r^2 - \beta r}\end{array} \qquad (14)$$

The substitution of Eqs. (12), (13) and (14) into Eq. (8) and dividing by $e^{-\alpha r^2 - \beta r}$ gives

$$F''(\mathrm{r}) + \left[-4\alpha r - 2\beta + \dfrac{2}{r}\right] F'(\mathrm{r}) + \left[\begin{array}{l}(4\alpha^2 + S)r^2 + (4\alpha\beta - B)r \\ + (P - 2\beta)\dfrac{1}{r} - \dfrac{L(L+1)}{r^2} + (\varepsilon + \beta^2 - 6\alpha)\end{array}\right] F(r) = 0 \qquad (15)$$

The function $F(r)$ is considered as a series of the form

$$F(r) = \sum_{n=0}^{\infty} a_n r^{2n+L} \qquad (16)$$

Taking the first and second derivatives of Eq. (16) we obtain,

$$F'(r) = \sum_{n=0}^{\infty}(2n+L)a_n r^{2n+L-1} \qquad (17)$$

$$F''(r) = \sum_{n=0}^{\infty}(2n+L)(2n+L-1)a_n r^{2n+L-2} \qquad (18)$$

We substitute Eqs. (16), (17) and (18) into Eq. (15) and obtain

$$\sum_{n=0}^{\infty}(2n+L)(2n+L-1)a_n r^{2n+L-2} + \left[-4\alpha r - 2\beta + \frac{2}{r}\right]\sum_{n=0}^{\infty}(2n+L)a_n r^{2n+L-1}$$
$$+\left[(4\alpha^2+S)r^2 + (4\alpha\beta-B)r + (P-2\beta)\frac{1}{r} - \frac{L(L+1)}{r^2} + (\varepsilon+\beta^2-6\alpha)\right]\sum_{n=0}^{\infty}a_n r^{2n+L} = 0 \qquad (19)$$

By collecting powers of $r$ in Eq. (19) we have

$$\sum_{n=0}^{\infty}a_n\begin{cases}\left[(2n+L)(2n+L-1)+2(2n+L)-L(L+1)\right]r^{2n+L-2}\\ +\left[-2\beta(2n+L)+(P-2\beta)\right]r^{2n+L-1}\\ +\left[-4\alpha(2n+L)+\varepsilon+\beta^2-6\alpha\right]r^{2n+L}\\ +\left[4\alpha\beta-Q\right]r^{2n+L+1}+\left[4\alpha^2+S\right]r^{2n+L+2}\end{cases} = 0 \qquad (20)$$

Equation (20) is linearly independent implying that each of the terms is separately equal to Zero, noting that $r$ is a non-zero function; therefore, it is the coefficient of $r$ that is zero. With this, we obtain the relation for each of the terms.

$$(2n+L)(2n+L-1)+2(2n+L)-L(L+1) = 0 \qquad (21)$$

$$-2\beta(2n+L)+P-2\beta = 0 \qquad (22)$$

$$-4\alpha(2n+L)+\varepsilon+\beta^2-6\alpha = 0 \qquad (23)$$

$$4\alpha\beta - Q = 0 \qquad (24)$$

$$4\alpha^2 + S = 0 \qquad (25)$$

From Eq. (22) we have

$$\beta = \frac{P}{4n+2L+2} \qquad (26)$$

From Eq. (25) we have

$$\alpha = \frac{\sqrt{-S}}{2} \qquad (27)$$

We obtain the energy equation using Eq. (23) and have

$$\varepsilon = 2\alpha(4n + 2L + 3) - \beta^2 \tag{28}$$

Substituting Eqs. (6), (9), (11), (26) and (27) into Eq. (28) we obtain the energy eigenvalues of HHP as,

$$E_{nl} = \sqrt{\frac{-\hbar^2 A_2 m_D^3(T)}{12\mu}}\left(4n + 2 + \sqrt{(2l+1)^2}\right)$$
$$-\frac{2\mu}{\hbar^2}\left(A_1 - A_2 + \frac{A_0}{m_D(T)}\right)^2\left(4n + 1 + \sqrt{(2l+1)^2}\right)^{-2} + \frac{A_0}{2} - A_2 m_D(T) \tag{29}$$

Special cases

1. Setting $A_0 = 0$ in Eq. (29) we obtain energy equation for Hellmann potential

$$E_{nl} = \sqrt{\frac{-\hbar^2 A_2 m_D^3(T)}{12\mu}}\left(4n + 2 + \sqrt{(2l+1)^2}\right)$$
$$-\frac{2\mu}{\hbar^2}\left(A_1 - A_2 + \frac{A_0}{m_D(T)}\right)^2\left(4n + 1 + \sqrt{(2l+1)^2}\right)^{-2} - A_2 m_D(T) \tag{30}$$

2. Setting $A_1 = A_2 = 0$ in Eq. (29) we obtain energy equation for Hulthén potential

$$E_{nl} = -\frac{2\mu}{\hbar^2}\left(+\frac{A_0}{m_D(T)}\right)^2\left(4n + 1 + \sqrt{(2l+1)^2}\right)^{-2} + \frac{A_0}{2} \tag{31}$$

3. Setting $A_0 = A_2 = 0$ in Eq. (29) we obtain energy equation for Coulomb potential

$$E_{nl} = -\frac{2\mu A_1^2}{\hbar^2}\left(4n + 1 + \sqrt{(2l+1)^2}\right)^{-2} \tag{32}$$

4. Setting $A_0 = A_1 = 0$ in Eq. (29) we obtain energy equation for Yukawa potential

$$E_{nl} = \sqrt{\frac{-\hbar^2 A_2 m_D^3(T)}{12\mu}}\left(4n + 2 + \sqrt{(2l+1)^2}\right)$$
$$-\frac{2\mu}{\hbar^2}\left(-A_2 + \frac{A_0}{m_D(T)}\right)^2\left(4n + 1 + \sqrt{(2l+1)^2}\right)^{-2} - A_2 m_D(T) \tag{33}$$

We test for the accuracy of the predicted results, using a Chi square function (Ali et al., 2020)

$$\chi^2 = \frac{1}{k}\sum_{i=1}^{k}\frac{\left(M_i^{Exp.} - M_i^{Theo.}\right)}{\Delta_i} \tag{34}$$

where $k$ runs over selected samples of heavy mesons, $M_i^{exp.}$ is the experimental mass of heavy mesons, while $M_i^{Th}$ is the corresponding theoretical prediction. The $\Delta_i$ quantity is experimental uncertainty of the masses. Intuitively, $\Delta_i$ should be one.

## RESULTS AND DISCUSSION

The mass spectra of the heavy mesons such as charmonium and bottomonium that have the quark and antiquark flavor is calculated by applying the following relation (Inyang, et al.,2021).

$$M = 2m + E_{nl}, \tag{35}$$

where $m$ is heavy quark mass, and $E_{nl}$ is energy eigenvalues.

By substituting Eq. (29) into Eq. (35) we obtain the mass spectra for HHP as:

$$M = 2m + \sqrt{\frac{-\hbar^2 A_2 m_D^3(T)}{12\mu}\left(4n + 2 + \sqrt{(2l+1)^2}\right)} \\ -\frac{2\mu}{\hbar^2}\left(A_1 - A_2 + \frac{A_0}{m_D(T)}\right)^2 \left(4n + 1 + \sqrt{(2l+1)^2}\right)^{-2} + \frac{A_0}{2} - A_2 m_D(T) \tag{36}$$

**Table 1.** Mass spectra of charmonium in (GeV) ($m_c$ =1.209 GeV, $\mu = 0.6045$ GeV, $A_0 = 1.422$ GeV, $A_1 = 2.949$ GeV, $A_2 = -0.009$ GeV, $m_D(T) = 1.52$ GeV, $\hbar = 1$)

| State | Present work | Abu-Shady, 2016 | Ciftci,and Kisoglu,2018 | Tanabashi et al., 2018 |
|---|---|---|---|---|
| 1S | 3.096 | 3.096 | 3.096 | 3.096 |
| 2S | 3.686 | 3.686 | 3.672 | 3.686 |
| 1P | 3.525 | 3.255 | 3.521 | 3.525 |
| 2P | 3.772 | 3.779 | 3.951 | 3.773 |
| 3S | 4.040 | 4.040 | 4.085 | 4.040 |
| 4S | 4.263 | 4.269 | 4.433 | 4.263 |
| 1D | 3.770 | 3.504 | 3.800 | 3.770 |
| 2D | 4.159 | - | - | 4.159 |
| 1F | 3.874 | - | - | - |

**Table 2.** Mass spectra of bottomonium in (GeV) ($m_b$ =4.823 GeV, $\mu = 2.4115$ GeV, $A_0 = -0.323$ GeV, $A_1 = 2.110$ GeV, $A_2 = -0.031$ GeV, $m_D(T) = 1.52$ GeV, $\hbar = 1$)

| State | Present work | Abu-Shady, 2016 | Ciftci,and Kisoglu,2018 | Tanabashi et al., 2018 |
|---|---|---|---|---|
| 1S | 9.460 | 9.460 | 9.462 | 9.460 |
| 2S | 10.023 | 10.023 | 10.027 | 10.023 |
| 1P | 9.898 | 9.619 | 9.963 | 9.899 |
| 2P | 10.256 | 10.114 | 10.299 | 10.260 |
| 3S | 10.355 | 10.355 | 10.361 | 10.355 |

| State | | | | |
|---|---|---|---|---|
| 4S | 10.580 | 10.567 | 10.624 | 10.580 |
| 1D | 10.164 | 9.864 | 10.209 | 10.164 |
| 2D | 10.306 | - | - | - |
| 1F | 10.209 | - | - | - |

## DISCUSSION

We predict the mass spectra of heavy mesons such as charmonium and bottomonium for different quantum states using Eq. (36). The free parameters of Eq. (36) were then obtained by solving two algebraic equations in the case of charmonium and bottomonium, respectively.

For bottomonium $b\bar{b}$ and charmonium $c\bar{c}$, we adopt the numerical values of these masses as $m_b = 4.823\,GeV$ and $m_c = 1.209\,GeV$ (Olive et al., 2014). Then, the corresponding reduced mass are $\mu_b = 2.4115\,GeV$ and $\mu_c = 0.6045\,GeV$, respectively. The experimental data were taken from (Tanabashi et al., 2018). We note that prediction of mass spectra of charmonium and bottomonium are in good agreement with experimental data and work of other researchers, as presented in Tables 1 and 2. In order to test for the accuracy of the predicted results, we used a Chi square function to determine the error between the experimental data and theoretical predicted values. The maximum error in comparison with the experimental data is found to be $0.034\,GeV$.

## CONCLUSION

In this study, we modeled the adopted HHP to interact in quark-antiquark system. The solutions of the Schrödinger equation for energy eigenvalues using the series expansion method were obtained. The energy eigenvalues was applied to predict heavy-meson masses of charmonium and bottomonium for different quantum states. The result agreed with experimental data with a maximum error of $0.034\,GeV$ and work of other researchers.

## ACKNOWLEDGEMENTS

Dr. Etido P.Inyang would like to thank Prof. A.N.Ikot for his support, that lead to the completion of my PhD research.